\documentclass[a4paper,10pt]{article}

\usepackage{amsmath}
\usepackage{amssymb}
\usepackage{graphicx}
\usepackage{psfrag}
\usepackage{color}
\usepackage{sistyle}
\usepackage{pifont}

\usepackage{setspace}
\DeclareMathOperator{\CNT}{\mathrm{CNT}}
\DeclareMathOperator{\HOPG}{\mathrm{HOPG}}
\DeclareMathOperator{\mica}{\mathrm{mica}}
\DeclareMathOperator{\peeling}{\mathrm{peeling}}

\definecolor{linkcolor}{rgb}{0,0,1}
\definecolor{red}{rgb}{0.6902,0.2941,0.2501}
\definecolor{blue}{rgb}{0.3059,0.4039,0.6157}

\newcommand\subfigletter[1]{}
\newcommand\subfigboxletter[1]{}

\def\d{\mathrm{d}}

\begin{document}

\title{\bf Carbon nanotubes adhesion and nanomechanical behavior from peeling force spectroscopy}

\author{Julien Buchoux$^{1}$, Ludovic Bellon$^{2}$, Sophie Marsaudon$^{1}$, Jean-Pierre Aim\'e$^{1}$\\
$^{1}$ Universit\'e Bordeaux 1 - CPMOH\\
351, Cours de la Lib\'eration, 33405 Talence Cedex, France
\\
$^{2}$
Universit\'e de Lyon
\\ Laboratoire de Physique, \'Ecole Normale Sup\'erieure de Lyon \\
CNRS UMR 5672 \\
46, All\'ee d'Italie, 69364 Lyon Cedex 07, France\\}

\maketitle

\begin{abstract}
Applications based on Single Walled Carbon Nanotube (SWNT) are good example of the great need to continuously develop metrology methods in the field of nanotechnology. Contact and interface properties are key parameters that determine the efficiency of SWNT functionalized nanomaterials and nanodevices. In this work we have taken advantage of a good control of the SWNT growth processes at an atomic force microscope (AFM) tip apex and the use of a low noise ($\SI{e-13}{m/\sqrt{Hz}}$) AFM to investigate the mechanical behavior of a SWNT touching a surface. By simultaneously recording static and dynamic properties of SWNT, we show that the contact corresponds to a peeling geometry, and extract quantities such as adhesion energy per unit length, curvature and bending rigidity of the nanotube. A complete picture of the local shape of the SWNT and its mechanical behavior is provided.
\end{abstract}

\vspace{10mm}

\noindent {\bf Keywords}

\noindent Thermal noise, carbon nanotube, adhesion, peeling

\vspace{5mm}

\noindent {\bf Notes}

\vfill

\noindent Corresponding authors :

-- Ludovic Bellon (Ludovic.Bellon@ens-lyon.fr, peeling experiments)

-- Sophie Marsaudon (s.marsaudon@cpmoh.u-bordeaux1.fr, nanotube synthesis).

\newpage

\section*{Introduction}

Since their discovery~\cite{Iijima-1991}, Single Walled Carbon Nanotube (SWNT) are at the origin of numerous creative works based on the conjugation of their exceptional electronic and mechanical properties. Their high aspect ratio with diameter in the $\SI{}{nm}$ range makes them perfect candidates as nano sensors. Ultimate detection at molecular level are envisioned with nanodevices in which carbon nanotubes (CNTs) are a central part: squid aiming at spin detection at the molecular level~\cite{Cleuziou-2006}, nanoscale resonators where coupling electronic charge transport with high frequencies CNT oscillation leads to highly sensitive mass detection~\cite{Lassagne-2009} or using exciton properties as to detect local variation of pH in biological environment~\cite{Cognet-2007}. Near field and nanoelectronic domains have motivated a wealth of attempts based on CNT as a central component~\cite{Avouris-2008,Anghel-2008}, while applications as original as fabrication of nanocomposite exhibiting thermal memory effect~\cite{Miaudet-2007} exploit its low density and high strength. Because the contact between the SWNT and the substrate or the surrounding medium always happens, the interface properties determine most of the nanodevice efficiency. For instance, contact between CNTs and electrodes might exhibit unwanted contact mechanical fluctuation leading to additional energy dissipation, while profile of polymer density surrounding CNT makes the interface properties a key parameter. Therefore, the conception and the fabrication of nanosystems need a companion development of metrology and methodology, a crucial step to develop efficient tools in nanotechnology.

In this work, we present experimental results from an original study of the SWNT mechanical behavior. Numerous works have been focused on mechanical properties of CNT~\cite{Kis-2004,Strus-2008,Marsaudon-2008,Kis-2008,Ishikawa-2008,Ishikawa-2009,Strus-2009-CST,Ke-2010}. However, accessing \emph{quantitative} information such as SWNT adhesion energy and mechanical properties of SWNT when brought in contact with a substrate is still a challenging experiment. In all existing experiments for example, measuring the adhesion energy per unit length of the CNT is either done in a very specific geometry (e.g. nanotube-nanotube interaction) ~\cite{Ke-2010,Benedict-1998,Chen-2003,Hertel-1998-PRB,Kis-2006} or using uncontrolled assumptions (e.g. CNT length in interaction)~\cite{Strus-2008,Ishikawa-2008,Ishikawa-2009,Strus-2009-CST}, leading to large incertitudes. In the present work, we take advantage of a good control of the SWNT growth processes at an Atomic Force Microscope (AFM) tip apex~\cite{Marty-2006} and the use of a high resolution AFM~\cite{Paolino-2009-Nanotech,Bellon-2010-HDR} to extract this information. The low level of noise ($\SI{e-13}{m/\sqrt{Hz}}$) affords simultaneous recording of static and dynamic measurements, i.e. force curves and thermal noise study, then giving a complete and precise picture of the SWNT mechanical behavior when the CNT partially touches a surface.

\section*{Experiments}

\subsection*{CNT Growth at the tip apex}

\begin{figure}
\begin{center}
\psfrag{1um}[Bc][Bc]{\small$\SI{1}{\micro m}$}
\psfrag{L}[Br][Br]{\small$L\ $}
\psfrag{dcost}[Br][Br]{\small$d\cos\theta_{0}$}
\psfrag{L-z}[Br][Br]{\small$L-z\ $}
\psfrag{Z}[Br][Br]{\small$Z$}
\psfrag{t}[Br][Br]{\small$\theta_{0}$}
\psfrag{Ra}[Br][Br]{\small$R_{a}$}
\includegraphics{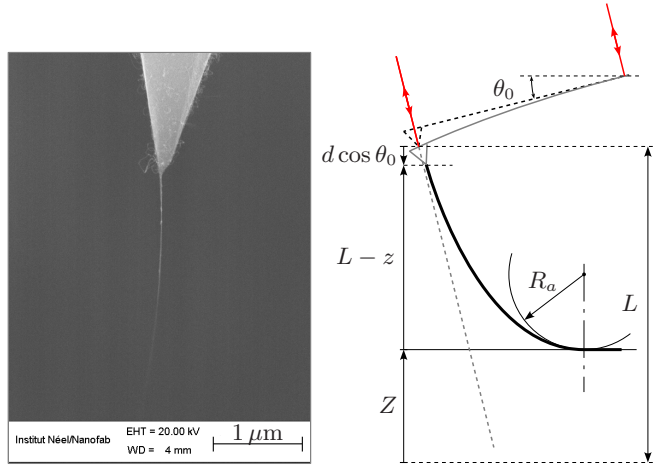}
\end{center}
\caption{\doublespacing Carbon nanotube tip. The SWNT is grown directly on the tip of an AFM cantilever, as shown on scanning electron micrograph. The length of the nanotube under study is $L\sim\SI{2}{\micro m}$. In the experiment, it is pressed against a substrate of graphite or mica, and the deflexion $d$ of the cantilever is measured with a differential interferometer. The optical path difference between the sensing beam, reflecting on the cantilever above the tip, and the reference beam on its base, is twice the deflexion $d$. The nanotube compression $z$ is inferred from the calibrated measurements of $d$ and $Z$ (sample vertical position), taking into account the common $\theta_{0}=\ang{15}$ angle between the force sensor and the sample. In the adsorbed state, balance between curvature and adsorption sets the radius of curvature $R_{a}$ at the last point of the free standing part of the nanotube.} 
\label{Fig:CNTtip}
\end{figure}

The growth of CNT at the tip apex uses the Hot Filament assisted Chemical Vapor Deposition (HFCVD) method~\cite{Marty-2006}. Several parameters are of importance to control SWNT growth, in particular it has been shown that  cobalt catalyst thickness is a critical parameter. Cobalt layer thickness superior to $\SI{9}{nm}$ leads to the growth of too many nanotubes at the tip apex, while for too thin one ($<\SI{4}{nm}$), the yield of SWNT growth at the apex of the Si tip becomes negligible. With an optimal catalyst thickness of the order of $(5-8)\,\SI{}{nm}$, the yield of production of a unique SWNT bundle at the apex of a commercial Si (similar to fig.~\ref{Fig:CNTtip}) tip is of about $\SI{30}{\%}$.

HRTEM observations demonstrate the formation of predominantly SWNT and double walled CNT, with diameters mostly found to be between $(\num{1.2}-\num{2.1})\, \SI{}{nm} \pm \SI{0.3}{nm}$~\cite{Fiawoo-2010}. Raman spectroscopy studies are also strong indications of the excellent crystalline property and purity of the grown CNT~\cite{Raman}. Therefore, HFCVD technique appears to be powerful to grow highly crystalline and pure SWNT at the apex on Si tips by taking advantage of the catalytic properties of a thin cobalt layer.  A scanning electron micrograph of the nanotube used in the experiment described in this paper is presented in fig.~\ref{Fig:CNTtip}. It is about $\SI{2}{\micro m}$ long, and well aligned with the tip. 

\subsection*{Peeling experiments}

\begin{figure}
\begin{center}
	\input{Figures/tps_freq_ab.tex}\includegraphics{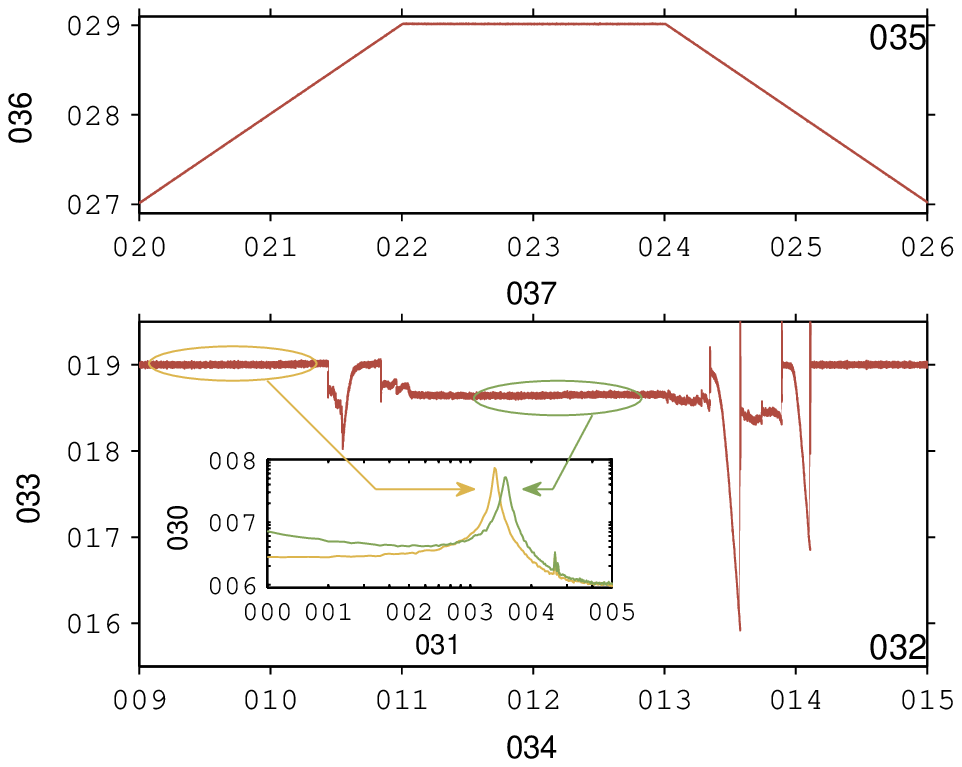}
	\input{Figures/tps_freq_c.tex}\includegraphics{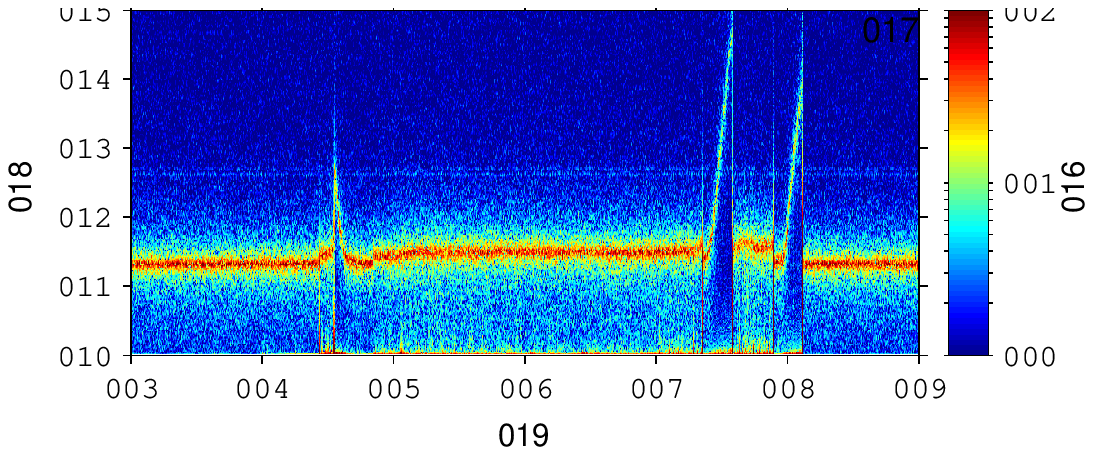}
\end{center}
\caption{\doublespacing Time frequency analysis of the deflection. Two top graphics present the time trace of the substrate position $Z$ and cantilever deflexion $d$ during an approach-retract cycle, corresponding to the force-compression curve of figure \ref{Fig:F&kCNT}. In the inset, a power spectrum density (PSD) of the deflexion signal is shown before (yellow) and during (green) contact: thermal noise excites the first resonance of the mechanical oscillator composed by the AFM cantilever and the CNT connecting the surface and the AFM tip. Before contact, the fit of the PSD with a simple harmonic oscillator model (dashed line) leads to the cantilever stiffness $k$. The dynamic stiffness $k_{\CNT}$ of the nanotube in contact can be computed from the observed frequency shift of the resonance. We generalize this technique with a time frequency analysis: every $\SI{5}{ms}$, we compute a PSD of the deflexion and plot the result in the color coded spectrogram of the bottom graphic. We extract from this plot the time evolution of the resonance frequency, and thus of the dynamic stiffness.}
\label{Fig:tpsfreq}
\end{figure}

In the experiment, we press the CNT against a flat surface of graphite or mica, and record the deflexion $d$ of the AFM cantilever as a function of the sample vertical position $Z$. The translation of the substrate is performed with a piezo translation platform operated in closed loop (PI P-527.3), featuring an accuracy of $\SI{0.3}{nm}$ rms. The measurement of the deflexion $d$ is performed with a home made interferometric deflection sensor~\cite{Paolino-2009-Nanotech,Paolino-2009-JAP,Bellon-2010-HDR}, inspired by the original design of Schonenberger~\cite{Schonenberger-1989} with a quadrature phase detection technique~\cite{Bellon-2002-OptCom}: the interferences between the reference laser beam reflecting on the base of the cantilever and the sensing laser beam on the free end of the cantilever (see figure \ref{Fig:CNTtip}) directly gives a measurement of $d$, with very high accuracy. The intrinsic background noise of our detector is only $\SI{e-13}{m/\sqrt{Hz}}$ for the cantilevers used in this experiment (see inset in figure \ref{Fig:tpsfreq}). Beyond this very low noise, one advantage of the technique is that it offers a calibrated measurement of the deflection, without conversion factor from Volt to meter as in the standard optical lever technique common in AFM. $Z$ and $d$ being both calibrated, we can therefore compute at any time the CNT compression $z=Z-d\cos(\theta_{0})$, and set the origin of $z$ at the last contact between the nanotube and the surface. In this formula, $\cos(\theta_{0})$ accounts for the $\theta_{0}=\ang{15}$ inclination of the AFM cantilever with the substrate.

We calibrate the spring constant $k$ of the cantilever with a thermal noise measurement far from the sample~\cite{Paolino-2009-JAP,Butt-1995}: the thermal excitation operates like a random force (white noise) on the cantilever, and we measure the resulting power spectrum density (PSD) of deflexion fluctuation. As illustrated by the inset of figure \ref{Fig:tpsfreq}, the PSD of the first resonance of the cantilever is well described by a simple harmonic oscillator model. From this fit, we determine the dynamic spring constant $k_{1}$ of the first mode of the cantilever: $k_{1}=(84 \pm 5).\SI{e-3}{N/m}$. The static stiffness $k$ is deduced from the dynamic one $k_{1}$ with a small correction coefficient computed for an Euler Bernoulli description of the cantilever~\cite{Butt-1995}: $k=0.97k_{1}=(81 \pm 5).\SI{e-3}{N/m}$. In quasi-static operation, the vertical force acting on the nanotube is computed by $F=kd/\cos(\theta_{0})$.

All signal are acquired at $\SI{200}{kHz}$ with high resolution acquisition cards (NI-4462) to determine the force compression curves $F$ vs $z$ when cycling the CNT against the substrate at low ramping speed (typically $\SI{500}{nm/s}$). Due to the finite stiffness of the cantilever, a mechanical instability occurring with attractive forces prevents continuous operation in equilibrium, and part of the force compression curves $F(z)$ cannot be accessed during the approach-retraction cycle. To exclude data that do not correspond to quasi-static operation of the cantilever, we discard any point presenting a deflexion speed $\d d/\d t$ greater than 4 standard deviations of its equilibrium fluctuations. 

If the ramp is sufficiently slow, we stay long enough around any compression $z$  to measure a spectrum of thermal noise driven fluctuations of deflexion. The force acting on the AFM tip is no longer due to the deflected cantilever alone, since the nanotube touching the surface has to be considered as well. The mechanical oscillator (cantilever first mode) experience an effective stiffness $k_{1}+k_{\CNT}$, shifting its resonance frequency from $f_{0}$ to $f_{0}+\Delta f$, as illustrated in the inset of figure \ref{Fig:tpsfreq}. In first approximation, the dynamic stiffness of the cantilever (around the resonance frequency of the oscillator) can be computed by~\cite{Buchoux-2009}:
\begin{equation}
k_{\CNT}=k_{1}\left[\left(1+\frac{\Delta f}{f_{0}}\right)^{2}-1\right] \label{kCNT}
\end{equation} 
We perform a time-frequency analysis of the deflexion signal, to access at each time to the PSD of the thermal noise driven deflexion. As long as the quasi-static approximation is valid and the resonance has a high enough quality factor, the maximum of the spectrum directly gives $\Delta f$, so we can use equation \ref{kCNT} to estimate $k_{\CNT}$. Figure \ref{Fig:tpsfreq} present a spectrogram of the deflexion signal during an approach-retract cycle. Each spectrum has been computed in a $\SI{5}{ms}$ time window, corresponding to a $\SI{2.5}{nm}$ translation of the sample. The thermal noise excitation is clearly strong enough to determine the resonance frequency shift, and thus $k_{\CNT}$. As $k_{\CNT}$ is inferred in the $\SI{10}{kHz}$ frequency range, it measures the dynamic stiffness of the nanotube/substrate system.

\begin{figure}
\begin{center}
\input{Figures/F-HOPG-mica.tex}\includegraphics{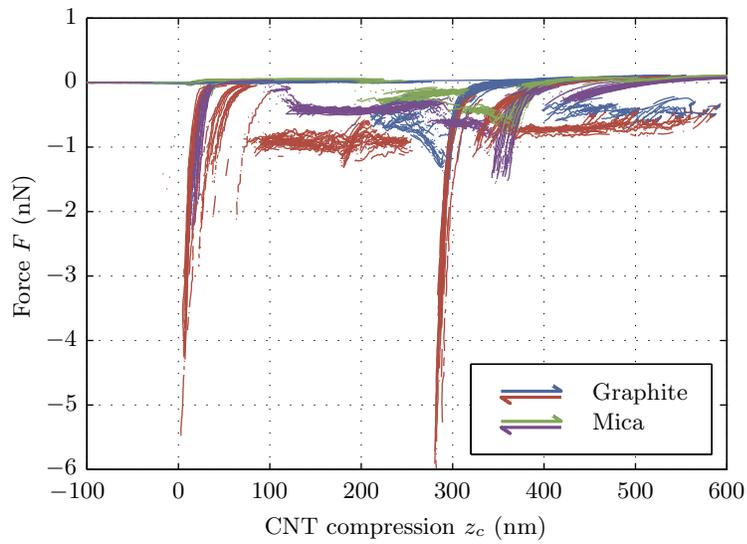}
\end{center}
\caption{\doublespacing Comparison of force-compression curves for graphite and mica: 40 independent measurements are plotted for each substrate, half with a ramping speed $\dot{Z}=\SI{0.5}{\micro m/s}$ and half with $\dot{Z}=\SI{5}{\micro m/s}$. The reproducibility of force plateaux is excellent, and characteristic of the nature of the sample.}
\label{Fig:F-HOPG&mica}
\end{figure}

Force-compression curves $F(z)$ and dynamic stiffness $k_{\CNT}(z)$ measured on graphite and mica surfaces are reported in figure \ref{Fig:F&kCNT}  and \ref{Fig:F&kCNT_mica} respectively. Our measurement device is fully calibrated for all observables, these are therefore quantitative measurements. An hysteresis between approach and retraction can be noticed, with anyway a perfect overlap of both force-compression profiles in some part of the curves. The interaction is everywhere attractive ($F<0$) or marginally repulsive, hinting at adhesion to be the most pertinent process to consider. Except for the two pronounced negative peaks during retraction, the force mainly presents two plateaux, one close to zero (principally during approach), the other around $\SI{-1}{nN}$ for graphite and $\SI{-0.4}{nN}$ for mica (principally during retraction). The very same characteristics can be noted on the stiffness curves, with diverging values of $k_{\CNT}$ connected to force accidents, and plateaux around zero and $\SI{0.4}{N/m}$ for graphite, $\SI{0.1}{N/m}$ for mica. As illustrated in figure \ref{Fig:F-HOPG&mica}, this behavior is very robust and the measurement is reproducible for different landing position of the substrate, ramping speeds (from $\SI{500}{nm/s}$ to $\SI{5}{\micro m/s}$), surrounding atmosphere (air or dry nitrogen) and for the two surfaces of graphite and mica. Values of the force and stiffness plateau are about half and one third on mica compare to that on graphite. 

\section*{Discussion}

To interpret those observations, let us model the nanotube by an elastic line, incompressible along its axis~\cite{Strus-2008,Oyharcabal-2005}. The shape of the force-compression curves, with two similar patterns (force plateaux, jumps and divergences) occurring reproducibly at about $\SI{300}{nm}$ distance for that specific nanotube, suggests that it is not mechanically integer on its whole $\SI{2}{\micro m}$ length. On other nanotubes (data not presented), similar reproducible patterns can be observed, with distances from few tens of nanometers up to $\SI{400}{nm}$. Stick-slip phenomenon can be discarded as the observed behavior is independent on ramping speed. A sound hypothesis is that the nanotube is composed of several ideal segments linked by defects presenting higher flexibility (kink like defects for example). The equilibrium shape of each segment can thus be described by the Elastica~\cite{Oyharcabal-2005}: 
\begin{equation} \label{eq:elastica}
EI \frac{\d^{2}\theta}{\d s^{2}} = EI \theta'' = F\sin{\theta}
\end{equation}
where $s$ is the arc length along the line, $\theta(s)$ the angle between the local slope and the vertical direction (normal to the sample), $E$ and $I$ are the Young's modulus and quadratic moment of the CNT, and $F$ the vertical force. We suppose that the nanotube segments can freely slide horizontally (either on the substrate or at their suspension point), so that we will not consider any horizontal forces. 

The interaction between the SWNT and the substrate is due to short range Van der Waals potential, it thus rapidly vanishes when the nanotube is not in immediate proximity (a few nanotube diameter at most) with the surface~\cite{Oyharcabal-2005}. We will therefore model this attractive interaction by a simple energy of adhesion per unit length $E_{a}$. The boundary condition on the substrate can be either a torque free condition ($\theta'=0$), corresponding to only the very end of the segment being in contact with the surface (point contact state), or an adsorption condition, when a non zero length of the nanotube is in contact with the substrate (adsorbed state). In this case, the continuity of the slope of the elastic line thus implies $\theta=\pi/2$ for the last point of the free standing part of the nanotube. As soon as part of the nanotube is adsorbed, minimizing the energy of the system will tend to maximize the absorbed length. However, this process increases the bending of the free standing part of the nanotube and its the associated curvature energy. The local shape of the CNT is the result of the balance between adhesion and bending, leading the radius of curvature at the contact point to be~\cite{Obreimoff-1930,Seifert-1991}
\begin{equation} \label{eq:R_a}
\frac{1}{\theta'}=R_{a}=\sqrt{\frac{EI}{2E_{a}}}
\end{equation}
If the free standing part of the nanotube has a length large compared to $R_{a}$,  the local shape of the SWNT does not change much when it is being peeled from the substrate. The vertical displacement $\delta z$ needed to peel a small length $\delta l$ is in first approximation $\delta z \simeq \delta l$. As we are pulling with a force $F$, the work produced is $F \delta z$ while the energy released is $E_{a} \delta l$, leading to $F\simeq E_{a}$: \emph{peeling the nanotube results in a flat force-compression curve}. This is indeed what is observed in our experiments, the value of the plateaux giving direct access to the value of the energy of adhesion per unit length: $E_{a}^{\HOPG}\simeq\SI{0.98}{nJ/m}$ for graphite substrate, and $E_{a}^{\mica}\simeq\SI{0.42}{nJ/m}$ for mica (see table \ref{table:numvalues} for details). These values are in good agreement with the few results available in the literature~\cite{Strus-2008,Ke-2010,Benedict-1998,Chen-2003,Hertel-1998-PRB,Kis-2006,Girifalco-2000,Raman-private}, though our error bars are much smaller.

In order to close the description of the nanotube as an elastic line, let us briefly discuss the boundary condition at the suspension point. When the nanotube is adsorbed, the torque applied at this point is small in the hypothesis of a high flexibility of this defect linking the segment under consideration to the upper part of the nanotube. We can thus use a torque free condition ($\theta'=0$) in this case. However, when the nanotube is not adsorbed, this boundary condition is not reasonable any more, since it would lead to a unrealistic straight shape for the considered segment. In this case, we will simply use a clamped hypothesis, with $\theta\sim\theta_{0}$ (where $\theta_{0}=\ang{15}$ is the common inclination of the AFM cantilever with the substrate).

\begin{figure}
 \begin{center}
	\input{Figures/elastica-BCtorquefree.tex}\includegraphics{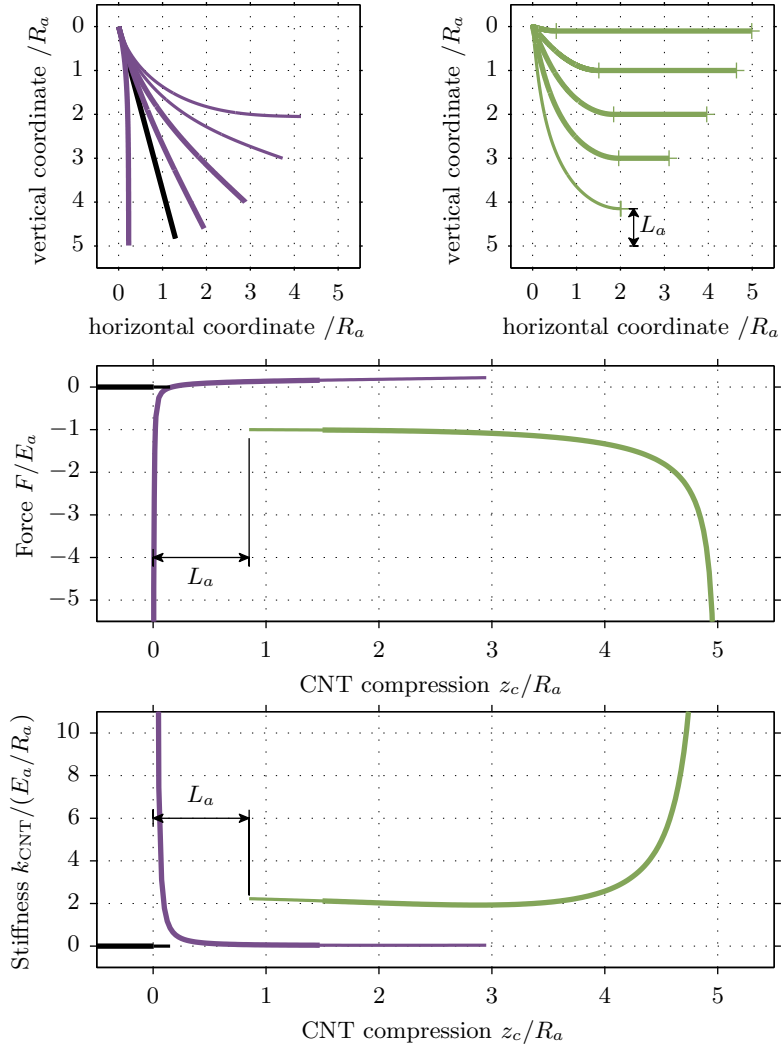}
\end{center}
\caption{\doublespacing Simulation of CNT compression for $L=5 R_{a}$. The nanotube is modeled as an elastic line incompressible along its axis, and different boundary conditions are considered whether it is adsorbed (green curves) or not (purple curves). Fundamental and metastable states are respectively drawn as thick and thin lines. The shape of the nanotube for various compression is plotted in the top graphic, while two bottom graphics present the vertical force $F$ and dynamic stiffness $k_{\CNT}$ as a function of compression $z$. During an approach-retraction cycle, the nanotube will first switch from its straight shape (black) to a weakly bended state (purple), which is (meta)stable till the CNT is tangent to the substrate. The nanotube will then adopt an adsorbed state (green), presenting force and dynamic stiffness plateau except for the highest compressions. Upon retraction, the nanotube can remain in this state as long as the adsorbed length (horizontal segment between ticks on top graphic) is not zero. It will then jump back to the weakly bended state, which will eventually correspond to a fully extended nanotube, presenting a diverging force before overcoming the adhesion energy of the CNT free end with the substrate. The arc length $L_{a}$ stored in the curved shape of the adsorbed nanotube and released when this state disappear can directly be read on the force-compression curve, and provides an estimation for $R_{a}$: $L_{a}=0.85R_{a}$.}
\label{Fig:elastica}
\end{figure}

We present in figure \ref{Fig:elastica} the force-compression curves numerically simulated for the model of an elastic line with a ratio $R_{a}/L=0.2$ where $L$ is the nanotube length (see \emph{Appendix} for details). During approach, the nanotube will first adopt a weakly bended shape, resulting in a small repulsive force. As the compression is increased, this state turns metastable, but may remain till the end of the nanotube gets tangent to the substrate. At this point, this branch of solution stops existing and part of the nanotube will be absorbed on the surface, on a force plateau close to the value of $E_{a}$. As the suspension point is brought closer to the surface, the force exhibit a divergence towards $-\infty$: let $L_{f}$ be the length of the free standing part of the nanotube, the distance between the surface and suspension point is of order $L_{f}^{2}/2 Ra$, so equalizing the work done by the force $F$ and the variation in adsorption energy leads to $F\sim - 2 E_{a} R_{a}/L_{f}$, diverging when $L_{f}$ tends to $0$.

\begin{figure}
\begin{center}
\psfrag{S1}[Bc][Bc]{\ding{192}}
\psfrag{S2}[Bc][Bc]{\ding{193}}
\psfrag{S3}[Bc][Bc]{\ding{194}}
\psfrag{S4}[Bc][Bc]{\ding{195}}
\psfrag{S5}[Bc][Bc]{\ding{196}}
\psfrag{S6}[Bc][Bc]{\ding{197}}
\psfrag{S7}[Bc][Bc]{\ding{198}}
\includegraphics{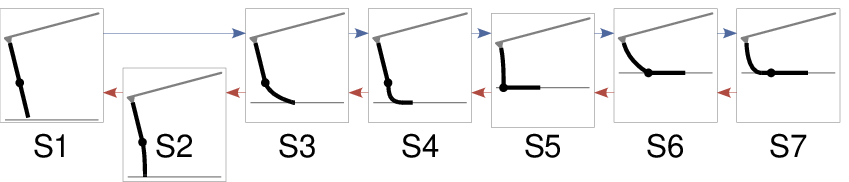}

\vspace{0.1cm}

\input{Figures/F-kCNT-HOPG.tex}\includegraphics{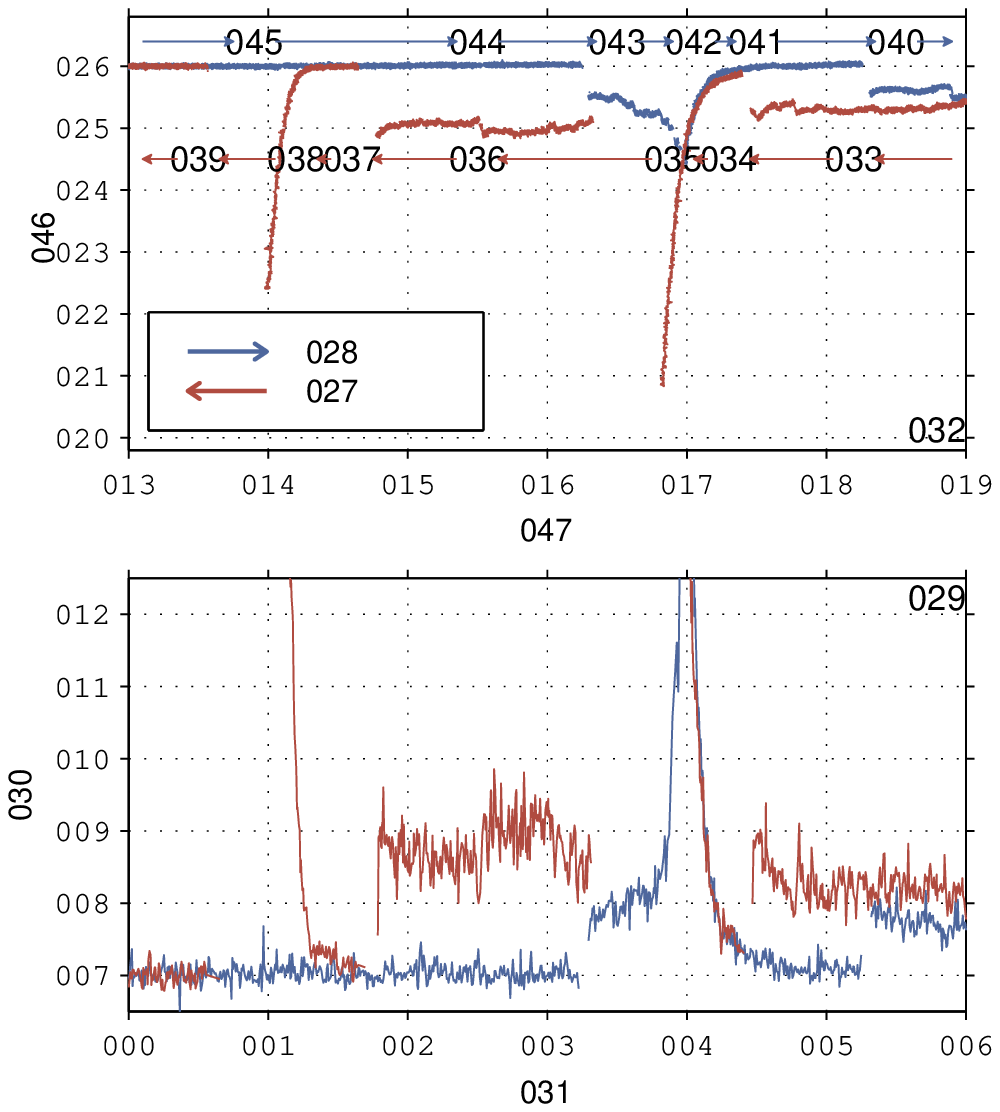}\vspace{-0.3cm}
\end{center}
\caption{\doublespacing Force $F$ and dynamic stiffness $k_{\CNT}$ of a nanotube as a function of its compression on a graphite substrate. A strong hysteresis, due to adhesion, can be noted between approach (blue) and retraction (red). Well defined plateaux of force (around $\SI{0.98}{nN}$) and stiffness (around $\SI{0.036}{N/m}$) allow one to estimate the energy of adhesion of the CNT on graphite ($E_{a}^{\HOPG}=\SI{0.98}{nJ/m}$), as well as the nanotube mechanical properties. The jumps and steep peaks of the curves are signature of transitions between point contact and adhesion shapes of various portions of the nanotube, as suggested by the scenario of numbered sketches. Off scale data for $k_{\CNT}$ climb up to $\SI{1}{N/m}$.}
\label{Fig:F&kCNT}
\end{figure}

\begin{figure}
\begin{center}
\input{Figures/F-kCNT-mica.tex}\includegraphics{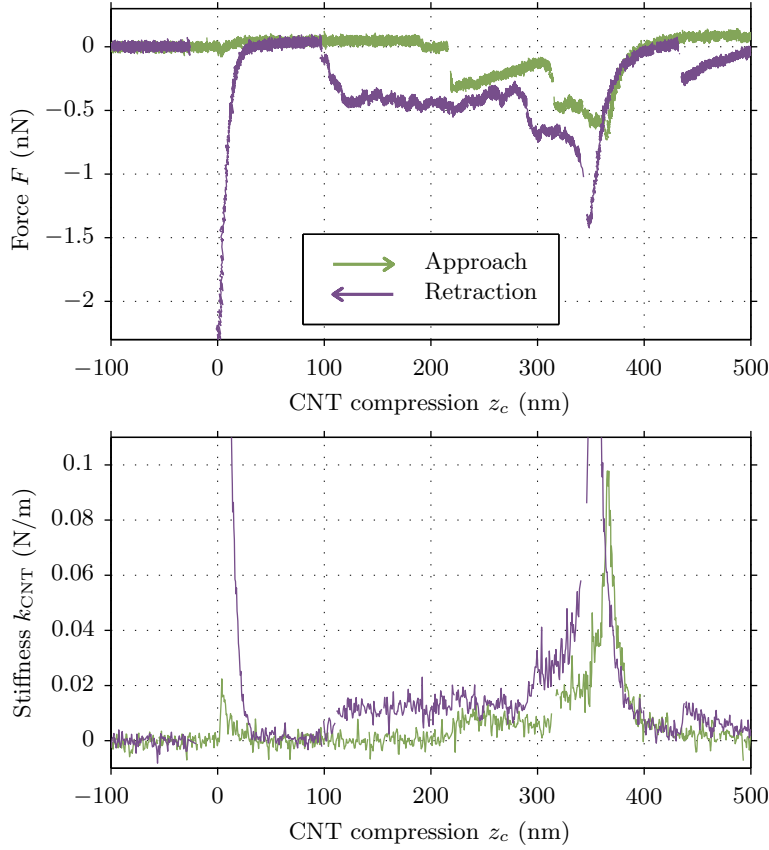}
\end{center}
\caption{\doublespacing Force $F$ and dynamic stiffness $k_{\CNT}$ of a nanotube as a function of its compression on a mica substrate. The curve is very similar to that of figure \ref{Fig:F&kCNT} with a graphite surface, except for the vertical scale: the energy of adhesion is estimated at $E_{a}^{\mica}=\SI{0.42}{nJ/m}$, about half of that with graphite. Similarly, the dynamic stiffness of a nanotube adsorbed on mica is one third of that on graphite: $k_{\CNT}^{\peeling}=\SI{0.013}{N/m}$.}
\label{Fig:F&kCNT_mica}
\end{figure}

During retraction, the adsorbed shape will remains stable or metastable as long as the absorbed length is non-zero, the force presenting the peeling plateau expected when the free standing length is larger than $R_{a}$. When the absorbed state disappear, the force will jump close to $0$ as the nanotube recover a weakly bended shape. The adhesion of the end point or next segment creates however a strong coupling to the surface. Further retraction will eventually lead to a fully extended nanotube (almost perpendicular to the surface), corresponding to very high negative forces. The connection with the substrate will finally break. The phenomenology expected from this model is very close to the experimental observations, as shown by the similarity between figures \ref{Fig:elastica} and \ref{Fig:F&kCNT} or \ref{Fig:F&kCNT_mica}.

Flat force-compression curves should lead to zero stiffness, since $\d F / \d z = 0$. However, it is clear on figures \ref{Fig:F&kCNT} and \ref{Fig:F&kCNT_mica} that $k_{\CNT}$ does not vanish when the force presents a plateau. This dynamic stiffness is measured in the $\SI{(10-50)}{kHz}$ range through thermal noise fluctuations. The discrepancy between $k_{\CNT}$ and $\d F / \d z$ can be explained by the simple assumption that adhesion is a slow process: in such a case, high frequency thermal fluctuations will only probe the response of the free standing part of the nanotube, the adsorbed length acting like a rigid clamping condition at fast time scales. The SWNT behaves as a nanomachine of finite stiffness linked to its local shape and its characteristic length $R_{a}$. In figure \ref{Fig:elastica}, we plot the dynamic stiffness computed under this hypothesis for the adsorbed elastic line. Just as for experimental data, it present a flat profile in correspondence to the static peeling force plateau. The value of this stiffness $k_{\CNT}^{\peeling}$ can be compared to $E_{a}/R_{a}$, the natural scale of the problem for the spring constant:  we get from our model
\begin{equation} \label{eq:kCNTpeeling}
k_{\CNT}^{\peeling}\simeq2\frac{E_{a}}{R_{a}}=\frac{EI}{R_{a}^{3}}
\end{equation}
Using mean value of the force and stiffness plateaux of figures \ref{Fig:F&kCNT} and \ref{Fig:F&kCNT_mica}, and equation \ref{eq:kCNTpeeling}, we estimate the radius  of curvature at the adhesion point for both substrates: $R_{a}^{\HOPG}=\SI{59}{nm}$ and $R_{a}^{\mica}=\SI{72}{nm}$  (see table \ref{table:numvalues} for details). 

The distance $L_{a}$ between the last adhesion point and the fully extended nanotube can also be used to estimate $R_{a}$: we read on figure \ref{Fig:elastica} $L_{a}=0.85 R_{a}$ for a ratio $R_{a}/L=0.2$. $L_{a}$ corresponds to the length stored in the curved shape and involved in the peeling process, that is restored as the peeling end up. In fact, as the last adhesion point is a metastable state (corresponding to a vanishing adsorbed length), the estimation we get through this observation gives an upper bound to $R_{a}$. This leads to $R_{a}^{\HOPG}\lesssim\SI{90}{nm}$ and $R_{a}^{\mica}\lesssim\SI{115}{nm}$, values that are coherent with the estimation through the dynamic stiffness.

From the definition of $R_{a}$ (equation \ref{eq:R_a}), its estimation together with the value of $E_{a}$ can be used to characterize the mechanical properties of the nanotube as we can compute the bending rigidity $EI$. This quantity should be independent of the substrate, and we find indeed a reasonable agreement (within 40\%, see table \ref{table:numvalues}) between the two measurements, with $EI\sim\SI{6e-24}{Jm}$. The bending rigidity is a characteristic mechanical property of the nanotube, linking its diameter $D_{\CNT}$ to its Young's modulus~\cite{Landau-1967}: $EI=\pi E  D_{\CNT}^{3} t_{\CNT}/8$, with $t_{\CNT}$ the thickness of the SWNT wall. Assuming $E=\SI{E12}{Pa}$ and $t_{\CNT}=\SI{0.34}{nm}$~\cite{Kis-2008}, we can compute a diameter $D_{\CNT}\sim\SI{3.5}{nm}$ for the nanotube segment probed in this experiment. This size is close to the expected diameter for our nanotubes.

\section*{Conclusion}

As a conclusion let us first summarize the main points we have developed in this article: we perform a series of experiments where a SWNT is pushed almost perpendicularly against a substrate of graphite or mica. We measure the quasi-static force as a function of the compression, but we can also access the dynamic stiffness using a analysis of thermal noise during this process. The most striking feature of these two observables is a plateau curve for a large range of compression, the values of which are substrate dependent. We use the Elastica to describe the shape of the nanotube, and a simple energy of adhesion per unit length $E_{a}$ to describe the interaction with the substrate. A natural length $R_{a}$ is defined, corresponding to the radius of curvature at the adsorption point when a non zero length of the nanotube is adsorbed on the substrate. $R_{a}$ results from an local equilibrium between curvature and adsorption. Comparison of experiment to a numerical integration of the problem demonstrate that the behavior to the nanotube is well described with these simple ingredients. The analysis of the experimental data naturally leads to the every quantity of interest in the problem (see table \ref{table:numvalues}): the force plateau is a direct measurement of the energy of adhesion per unit length $E_{a}$ for each substrate, and we easily determine $R_{a}$ from the dynamic stiffness plateau. Mechanical properties of the nanotube itself (its bending rigidity $EI$) can be extracted from those values, and prove to be independent of the substrate.

This work provides quantitative values on the adhesion energy between a SWNT and a substrate. The key point for the method to work is a weak rigidity with respect to adsorption: the radius of curvature at the adsorption point should be small compared to the length of the nanotube.  Further investigation should as well give numbers on gold or platinum substrate or any metallic surfaces. Therefore, the present experimental setup should help to design the most appropriate contact electrode for SWNT based nanodevices. Another issue we can address is the access to quantitative information on interface properties between polymer materials and carbon nanotubes. This application deals with the important field of designing composite polymer materials reinforced with CNT~\cite{Breuer-2004,Hussain-2006,Bal-2007}.

In adsorbed configuration, the SWNT acquires an equilibrium curvature shape with a bending elastic energy balancing the adhesion energy. A promising development is to further exploit this configuration as a highly sensitive mechanical nanomachine. Both the flat pulling force (in the $\SI{}{nN}$ range) and the spring constant (in the $\SI{E-2}{N/m}$ range) of this nanomachine are directly related to the energy of adhesion. Therefore, any change of the adhesion due to molecule surfactant or other perturbation, for instance local variation of pH in liquid environment, can be monitored with a great precision of 2 independent variables. Periodic perturbations can as well be detected. In particular if there are any characteristic frequencies governing the contact length fluctuation, for the frequencies of the perturbation that fall in that range a stochastic resonant process~\cite{Gammaitoni} can be called to amplify the detection sensitivity.

\appendix

\section*{Appendix : numerical simulation}

The equilibrium shape of the elastic line used to describe the nanotube is a solution of the Elastica (equation \ref{eq:elastica}), which usually doesn't have simple analytic solutions. We use Matlab to solve this ordinary differential equation using boundary conditions previously described:
\begin{itemize}
\item \emph{unabsorbed state}: clamped extremity at the suspension point ($\theta(s=0)=\theta_{0}=\ang{15}$) and torque free condition at nanotube end ($\theta'(s=L)=0$).
\item \emph{absorbed state}: torque free condition at the origin ($\theta'(s=0)=0$) and clamped extremity at the contact point with the substrate ($\theta(s=L_{f})=\pi/2$, where $L_{f}$ is the length of the free standing part of the nanotube above the surface).
\end{itemize}
The natural control parameter when solving the Elastica is the external force $F$, whereas the experimental control parameter is the nanotube compression $z=L-\int_{0}^{L_{f}}\cos(\theta)\d s$. This integral condition is not easy to handle directly, so we perform a shoot and adjust strategy to find for any $z$ the corresponding force $F$.  An additional step is required in the adsorbed state as the length of nanotube in contact is also a free parameter. We adjust this variable by a minimization of the total energy of the system $E_{c}-(L-L_{f})E_{a}$, where $E_{c}=EI/2 \int_{0}^{L_{f}}{\theta'}^{2} \d s$ is the curvature energy of the free standing part, and $(L-L_{f})E_{a}$ the energy of the adsorbed part of the CNT. When both states can exist for a given $z$, we compare their total energies to know which of the two is metastable.

\newpage

\bibliographystyle{unsrt}
\bibliography{CNTadhesion}

\newpage

\begin{table}[htdp]
\begin{center}
\begin{tabular}{|l|c|c|c|c|c|c|}
\hline
Substrate & $E_{a} / (\SI{}{nJ/m}) $ & $k_{\CNT}^{\peeling}/(\SI{}{N/m})$ & $R_{a}/(\SI{}{nm})$ & $L_{a}/(\SI{}{nm})$ & $EI/(\SI{e-24}{Jm})$ & $D_{\CNT}/(\SI{}{nm})$ \\
\hline
HOPG & $0.98\pm0.07$ & $0.036\pm0.007$ & $59\pm10$ & $85\pm15$ & $7.1\pm2.5$ & $3.7\pm0.4$ \\
\hline 
mica & $0.42\pm0.04$ & $0.013\pm0.003$ & $72\pm15$ & $115\pm15$ & $4.5\pm2.0$ & $3.2\pm0.5$ \\
\hline
\end{tabular}
\end{center}
\caption{\doublespacing Measured values for the adhesion energy per unit length $E_{a}$, dynamic spring constant plateau $k_{\CNT}^{\peeling}$, radius of curvature at adhesion point $R_{a}$, stored length in absorbed shape $L_{a}$, bending modulus of the nanotube $EI$ and estimated nanotube diameter $D_{\CNT}$ for two different substrates. Data correspond to mean values and standard deviations on the plateau of force and stiffness for compression $z$ in the $(80-220)\,\SI{}{nm}$ range for graphite, and $(120-280)\,\SI{}{nm}$ range for mica, except $L_{a}$ which is determined as indicated in figure \ref{Fig:elastica} using the 80 experiments displayed in figure \ref{Fig:F-HOPG&mica}.}
\label{table:numvalues}
\end{table}

\newpage	

\newpage

\newpage

\newpage

\newpage

\newpage

\end{document}